\documentclass[twocolumn,prl,superscriptaddress,aps,showpacs]{revtex4}

\usepackage{graphicx}
\usepackage{dcolumn}
\usepackage{amsmath}
\newlength{\figwidth}
\newlength{\figwidthb}
\begin{document}


\title{Comparison of Resonant Inelastic X-Ray Scattering Spectra and Dielectric Loss Functions in Copper Oxides}
\author{Jungho Kim}
\affiliation{Department of Physics, University of Toronto, Toronto,
Ontario M5S~1A7, Canada}
\author{D. S. Ellis}
\affiliation{Department of Physics, University of Toronto, Toronto,
Ontario M5S~1A7, Canada}
\author{H. Zhang}
\affiliation{Department of Physics, University of Toronto, Toronto,
Ontario M5S~1A7, Canada}
\author{J. P. Hill}
\affiliation{Department of Condensed Matter Physics and Materials
Science, Brookhaven National Laboratory, Upton, New York 11973, USA}
\author{F. C. Chou}
\thanks{Present Address: Center for Condensed Matter Sicences,
National Taiwan University, Taipei 10717, Taiwan.}
\affiliation{Center for Materials Science and Engineering, MIT,
Cambridge, Massachusetts 02139, USA}
\author{T. Gog}
\affiliation{XOR, Advanced Photon Source, Argonne National
Laboratory, Argonne, Illinois 60439}
\author{D. Casa}
\affiliation{XOR, Advanced Photon Source, Argonne National
Laboratory, Argonne, Illinois 60439}
\author{Young-June Kim}
\email{yjkim@physics.utoronto.ca} \affiliation{Department of
Physics, University of Toronto, Toronto, Ontario M5S~1A7, Canada}

\date{\today}

\begin{abstract}
We report empirical comparisons of Cu K-edge indirect resonant
inelastic x-ray scattering (RIXS) spectra, taken at the Brillouin
zone center, with optical dielectric loss functions measured in a
number of copper oxides. The RIXS data are obtained for
Bi$_2$CuO$_4$, CuGeO$_3$, Sr$_2$Cu$_3$O$_4$Cl$_2$, La$_2$CuO$_4$,
and Sr$_2$CuO$_2$Cl$_2$, and analyzed by considering both incident
and scattered photon resonances. An incident-energy-independent
response function is then extracted. The dielectric loss functions,
measured with spectroscopic ellipsometry, agree well with this RIXS
response, especially in Bi$_2$CuO$_4$ and CuGeO$_3$.
\end{abstract}


\maketitle

Since the pioneering study of NiO by Kao $et~al.$ \cite{Kao96},
indirect resonant inelastic x-ray scattering (RIXS)
\cite{vandenBrink06,Ament07} in the hard x-ray regime has been
regarded as a promising momentum-resolved spectroscopic tool for
investigating charge excitations in solids
\cite{Kotani01,Platzman98,Hill98}. A variety of systems, including
nickelates \cite{Collart06,Wakimoto08}, manganites
\cite{Inami03,Grenier05} and cuprates
\cite{Abbamonte99,Hasan00,LCO-PRL,112-RIXS,Ishii05a,Ishii05b,Suga05,Lu05,Hill08},
have been studied by K-edge RIXS, revealing new and interesting
charge excitations. Despite these early experiments, further
developments of this promising technique have been hindered by the
lack of a systematic theoretical understanding. The RIXS
cross-section is complicated because the resonant contribution in
RIXS generally involves correlation between more than two particles
\cite{Devereaux07,Platzman98,vandenBrink06}. To date, theoretical
studies of the RIXS cross section have mostly been limited to
model-dependent calculations based on small clusters
\cite{Tsutsui99,Nomura05,Markiewicz06,Vernay08}. This is in contrast
with the cross-section for non-resonant inelastic x-ray scattering
(IXS) which measures the two-particle charge correlation function
\cite{Platzman73}. Although there have been a few attempts to
empirically relate the RIXS cross-section to a calculable
two-particle correlation function
\cite{Abbamonte99,112-RIXS,Grenier05,xsection,vandenBrink06,Ament07},
details of the incident and scattered photon resonances and the
nature of the response function measured by RIXS remain unclear.

In this context, recent theoretical studies by Ament and coworkers
are noteworthy \cite{vandenBrink06,Ament07}. Working in the limits
of a local, strong (or weak), and short-lived, core hole potential,
they were able to show that, in these limits, the RIXS cross section
can be factored into a resonant prefactor that depends on the
incident and scattered photon energies, and the dynamic structure
factor, $S(\textbf{q},\omega)$ \cite{Ament07}. This result has
important implications for the interpretation of RIXS spectra, since
this approach then suggests that with proper handling of the
prefactor, RIXS can be considered as a weak probe that measures
$S(\textbf{q},\omega)$. It is therefore important to test whether
this theory can be applied to real systems or not, and delineate any
necessary conditions for the applicability of Ref.~\cite{Ament07}.

In this Letter, we report a systematic comparison of the RIXS
spectra and the dielectric loss functions in various copper oxides:
Bi$_2$CuO$_4$, CuGeO$_3$, Sr$_2$Cu$_3$O$_4$Cl$_2$, La$_2$CuO$_4$,
and Sr$_2$CuO$_2$Cl$_2$. We measured the RIXS spectrum of each
sample at the $\Gamma$-position and extracted a response function
that does not depend on the incident energy. This response function
is compared with the optical dielectric loss function, measured
using spectroscopic ellipsometry on the same sample. Dielectric loss
function data from published electron energy loss spectroscopy
(EELS) studies are also used to augment the optical data. We show
that overall energy loss-dependence of the RIXS response function is
in good agreement with the optical dielectric loss function over a
wide energy range for all samples. In particular, the agreement is
excellent for the local systems Bi$_2$CuO$_4$ and CuGeO$_3$,
suggesting that the RIXS response is intimately related to the
dynamic structure factor and thus to the charge density-density
correlation function in these materials. On the other hand,
additional low-energy spectral features are observed for the
corner-sharing two dimensional copper oxides such as La$_2$CuO$_4$
and Sr$_2$CuO$_2$Cl$_2$, indicating that non-local nature of
intermediate states may play an important role in such compounds.


The Cu K-edge RIXS experiments were carried out at the 9IDB beamline
at the Advanced Photon Source. Channel-cut Si(444)/Si(333)
monochromators, various detector slits, and a diced Ge(733) analyzer
with 1 m radius of curvature were utilized. The data shown here are
obtained with low resolution (FWHM of the elastic line: $300-
400$~meV) unless otherwise specified. The single crystal samples of
Bi$_2$CuO$_4$, CuGeO$_3$, La$_2$CuO$_4$ were grown using the
traveling solvent floating zone method, while the
Sr$_2$Cu$_3$O$_4$Cl$_2$ and Sr$_2$CuO$_2$Cl$_2$ single crystals were
prepared using the CuO flux method. The experimental conditions for
each measurement are summarized in Table~I. All RIXS measurements
were carried out at the appropriate $\textbf{q}$=0 reciprocal space
positions. We have also studied the optical properties of the same
samples with spectroscopic ellipsometry using a VASE (Woollam)
ellipsometer. The real and imaginary part of the dielectric function
were obtained from 1~eV to 6.2~eV at room temperature \cite{pseudo}.


From the Eq.~(26) in Ref.~\cite{Ament07}, the RIXS intensity
$I(\omega, \omega_i)$ at {\bf q}=0 can be written as
\cite{spin_part}
\begin{eqnarray}
I \sim P(\omega, \omega_i)S(\omega)\delta(\omega-\omega_s+\omega_i),
\end{eqnarray}
where $\omega_i$ and $\omega_s$ are the incident and scattered
photon energies, respectively, and $\omega$=$\omega_i-\omega_s$ is
the energy loss. Here, $S(\omega) \equiv S(\bf{q}=0,\omega)$ is the
density response function at $\bf{q}=0$
\cite{Platzman98,Abbamonte99}.  Note that the resonance behavior
involving intermediate states is all contained in the prefactor
$P(\omega,
\omega_i)=[(\omega_s-\omega_{res})^2+\Gamma^{2}]^{-1}[(\omega_i-\omega_{res}-|U|)^2+\Gamma^2]^{-1}$,
which consists of two Lorenzian functions with damping (${\Gamma}$)
due to the finite lifetime of intermediate states. Each of these
Lorentzian functions represent resonant behavior of incident and
scattered photons, resonating at $\omega_{res}+|U|$ and
$\omega_{res}$, respectively, where $|U|$ is the local core hole
potential.

\begin{table}
\caption {Summary of the experimental conditions and fitting
parameters for the samples studied. Energies are in units of eV.
Note that the $\rm CuGeO_3$ data were taken at the 1D zone center of
the chain.} \label{table1}
\begin{ruledtabular}
\begin{tabular}{ccclllr}
Sample & Q (rlu) & polarization & $\omega_{res}$ & $\Gamma$ & $|U|$ & T (K) \\
\hline
$\rm Bi_2CuO_4$     & (0,0,3)     &$\epsilon \perp z$ & 8993.7 & 3.2 & 5.7 & 20 \\
$\rm CuGeO_3$       & (1.5,0,0)   &$\epsilon \sim \parallel z$\footnote{The polarization was along the crystallographic b-direction.} & 8985.7 & 2.5 & 4.3 & 300 \\
$\rm Sr_2Cu_3O_4Cl_2$ & (0,0,9)   &$\epsilon \perp z$ & 8989.0 & 3.5 & 4.5 & 300 \\
$\rm La_2CuO_4$     & (3,0,0)     &$\epsilon \parallel z$ & 8989.3 & 2.5 & 2.5 & 20 \\
$\rm Sr_2CuO_2Cl_2$ & (0,0,11)    &$\epsilon \perp z$ & 8992.4 & 3.0 & 2.6 & 300 \\
\end{tabular}
\end{ruledtabular}
\end{table}

We have analyzed our data by fitting all scans using Eq. (1),
leaving $\omega_{res}$, $\Gamma$, and the overall amplitude as free
parameters. The fitting parameters for each sample are listed in
Table~I. Note that the peak position of the K-edge x-ray absorption
spectrum ($\omega_{XAS}$) is taken to be the incident photon
resonance energy $\omega_{res}+|U|$. Since the incident photon
energy $\omega_i$ is fixed for each scan, the
$[(\omega_i-\omega_{XAS})^2+\Gamma^{2})]^{-1}$ factor becomes a
constant for each scan, leaving a single Lorentzian function in the
prefactor. This remaining Lorentzian function centered at
$\omega_{res}$ with width $\Gamma$ describes the scattered photon
resonance, which is necessary to obtain satisfactory fits
\cite{self}.

In order to illustrate the use of the scattered photon resonance,
the RIXS spectra of $\rm CuGeO_3$ obtained with four different
incident energies are plotted as a function of $\omega_{s}$ in Fig.
1(a). For each scan, there are two strong energy loss features
centered at $\omega=3.8$ eV and 6.5 eV. Although the energy-loss
positions of these features do not change from scan to scan, the
ratio between these two features varies quite a lot. The
$\omega=6.5$ eV peak is strongest when $\omega_i=8992$ eV, while the
$\omega=3.8$ eV peak is strongest when $\omega_i=8990$ eV. These
observations seem to suggest that the energy loss is a
material-specific property arising from $S(\omega)$, while the
intensity variation is due to the resonance prefactor. The scattered
photon resonant prefactor, shown as the Lorentzian envelope function
in Fig. 1(a), describes the intensity ratio variation between these
two peaks very well. In other words, by fitting all our data to
Eq.~(1), it is possible to extract an $\omega_i$-independent part of
the spectrum: $S(\omega)$.

%
%

\begin{figure}[t]
\centerline{\includegraphics[width=3in]{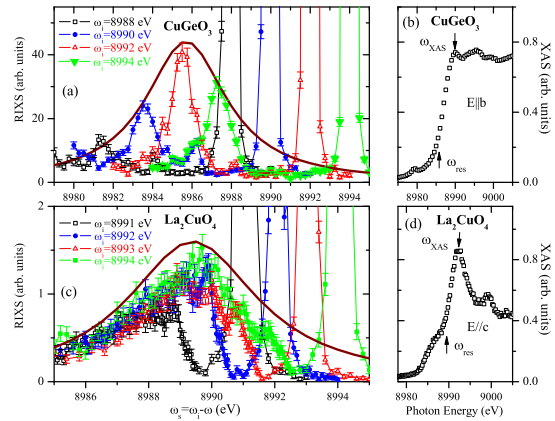}}
\vspace*{0.0cm}%
\caption{(Color online) (a) The K-edge RIXS spectra taken with four
incident photon energies as a function of the scattered photon
energy ($\omega_s$) for CuGeO$_3$. Note that the inelastic intensity
follows an envelope function (solid line), indicating the scattered
photon resonance. (b) The total fluorescence yield obtained near the
Cu K-edge for CuGeO$_3$. Two resonance energies, $\omega_{XAS}$ and
$\omega_{res}$, are noted. Similar plots for La$_2$CuO$_4$ are shown
in (c) and (d). The RIXS data for La$_2$CuO$_4$ are taken from
Ref.~\cite{LCO-PRL}.}\label{fig:resonance}
\end{figure}

%
%

\begin{figure}[t]
\centerline{\includegraphics[width=2.5in,angle=0]{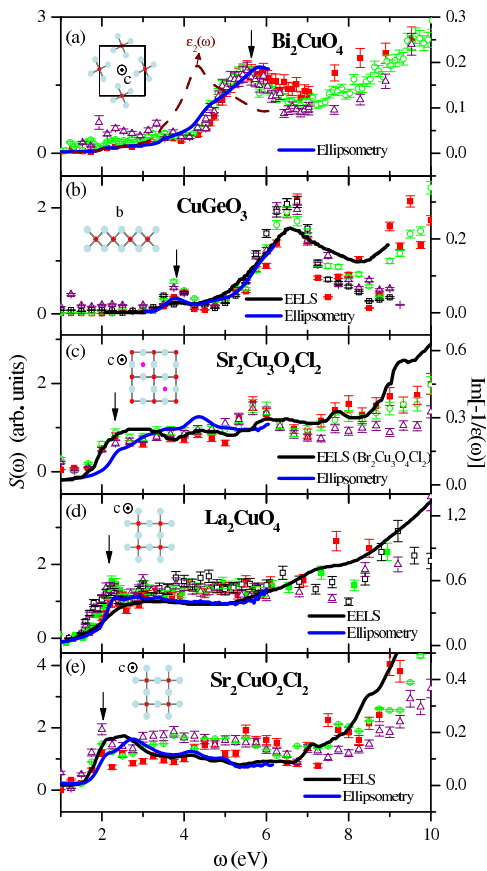}}
\vspace*{0.0cm}%
\caption{(Color online) Comparison of ${S}(\omega)$ with the
measured Im(-1/$\varepsilon(\omega)$) for various cuprates.
Different symbols are used to denote RIXS spectra obtained with
different incident energies. These incident photon energies are (a)
$8997.5 - 8999.5$ eV; (b) $8988 - 8994$ eV; (c) $8992 - 8994$ eV;
(d) $8991 - 8994$ eV; and (e) $8993 - 8995$ eV. The EELS data for
$\rm CuGeO_3$, $\rm La_2CuO_4$, and $\rm Sr_2CuO_2Cl_2$ are taken
from Refs.~\cite{Atzkern01}, \cite{Terauchi99}, and
\cite{Moskvin02}, respectively. Although there is no EELS spectrum
for $\rm Sr_2Cu_3O_4Cl_2$ available in the literature, that of $\rm
Ba_2Cu_3O_4Cl_2$ is shown \cite{Knauff96}.}\label{fig:compare}
\end{figure}

Similar analysis of the $\omega_i$ and $\omega_s$ dependence of the
spectra has also been carried out for the other samples. The data
for La$_2$CuO$_4$ are shown in Fig.~\ref{fig:resonance}(c)-(d).
Again the envelope function seems to describe the intensity
variation as a function of $\omega_i$ well. In
Fig.~\ref{fig:compare}(a)-(e), the resulting ${S}(\omega)$ spectra
are shown as a function of $\omega$. The elastic tail in the RIXS
raw data are fitted to a Lorentzian function and subtracted in this
plot. We find that, for a given material, the spectra obtained with
different incident energies collapse onto a single spectrum that
represents ${S}(\omega)$, as expected. We note that a similar
analysis was carried out in Refs.~\cite{Abbamonte99,Doring04,Lu06},
but with the assumption that the incident and scattered intermediate
states were degenerate ($U=0$) \cite{Abbamonte99,Doring04} or nearly
degenerate ($|U|=1$~eV) \cite{Lu06}. One could take the same
approach here, but we found that assuming two distinct intermediate
states for $\omega_i$ and $\omega_s$ is essential in describing all
the spectral features simultaneously. As discussed in
Ref.~\cite{Ament07}, these two resonances come from the intermediate
state energies separated by the core hole potential. Table~I lists
the values of $\omega_{res}$ obtained from our fitting and
$|U|=\omega_{XAS}-\omega_{res}$. It should be noted that $|U| <
\Gamma$ for $\rm La_2CuO_4$ and $\rm Sr_2CuO_2Cl_2$.

Next, we compare the extracted ${S}(\omega)$ with the measured
optical response functions. One of the key results in
Ref.~\cite{Ament07} is that apart from the prefactor, the RIXS
should measure $S(\omega)$, which captures the physics of the
valence electron system. Several studies have also argued that the
response function of RIXS corresponds to the dynamic structure
factor, $S(\bf{q},\omega)$, or equivalently the dielectric
loss-function, Im(-1/$\varepsilon(\omega)$)
\cite{Abbamonte99,112-RIXS,Grenier05,xsection,vandenBrink06,Ament07}.
Having obtained an incident-energy-independent RIXS response
function, we are now in a position to test this idea directly. To do
so, we obtained optical spectra from ellipsometry measurements on
the same samples used in the RIXS experiments. As a first
comparison, we plot both the imaginary part of complex dielectric
function, $\varepsilon_2(\omega)$, and Im(-1/$\varepsilon(\omega)$)
in Fig.~\ref{fig:compare}(a). It is clear that ${S}(\omega)$ is most
similar to Im(-1/$\varepsilon(\omega)$), but not to
$\varepsilon_2(\omega)$. In Fig.~\ref{fig:compare}, measured
Im(-1/$\varepsilon(\omega)$) spectra for the other systems are
overlaid with the ${S}(\omega)$ spectra. We also include the
available EELS spectra near $\textbf{q}=0$ from the literature
\cite{Atzkern01,Knauff96,Terauchi99,Moskvin02}. It is well known
that EELS at $\textbf{q}=0$ also measures
Im(-1/$\varepsilon(\omega)$). Note that the RIXS data are plotted
using an arbitrary vertical scaling in making these comparison.

Figure~\ref{fig:compare} shows that the extracted ${S}(\omega)$ is
in good overall agreement with the general shape of
Im(-1/$\varepsilon(\omega)$) for all five systems studied,
regardless of the detailed characteristics of the particular charge
excitations. In particular, Bi$_2$CuO$_4$ and CuGeO$_3$ show
excellent agreement in terms of the number of peaks and their
positions. In the other systems, however, it is not easy to see a
one-to-one correspondence with particular features, in part because
the spectra themselves are more complex with several features in the
low energy region. These will be discussed in more detail later.
Nevertheless, as indicated by the arrows in
Fig.~\ref{fig:compare}(c)-(e), the position of the insulating gap
and the increase in $S(\omega)$ above $7 \sim 8$~eV agree reasonably
well for the two data sets.

%
%

\begin{figure}[t]
\centerline{\includegraphics[width=2.2in,angle=0]{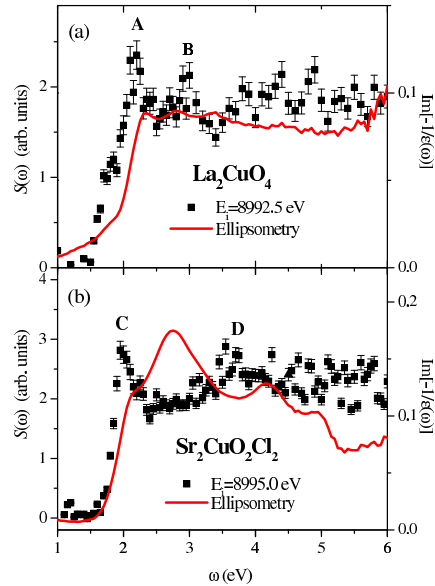}}
\caption{(Color online) The function ${S}(\omega)$ extracted from
high-resolution RIXS data and compared with
Im(-1/$\varepsilon(\omega)$) for (a) La$_2$CuO$_4$ and (b)
Sr$_2$CuO$_2$Cl$_2$.} \label{fig:detail}
\end{figure}

To make a detailed comparison for $\rm La_2CuO_4$ and $\rm
Sr_2CuO_2Cl_2$, we plot high-resolution ($\sim$130~meV) RIXS data
and ellipsometry data for these two materials in
Fig.~\ref{fig:detail} focusing on the low energy region. The
high-resolution ${S}(\omega)$ is obtained via the same procedure as
described above, with the same parameters. Features labeled A and C
in the two RIXS spectra correspond to the lowest energy charge
excitations, and presumably have the same origin as the charge gap
in Im(-1/$\varepsilon(\omega)$), yet their spectral shape appears
different for the two techniques. We also observe an additional
peak, B, around 3~eV, in $\rm La_2CuO_4$, which is not apparent in
the ellipsometry data. Similarly, feature D in Sr$_2$CuO$_2$Cl$_2$
is not observed in the ellipsometry, while strong features are
observed between C and D. In the case of La$_2$CuO$_4$,
${S}(\omega)$ has a small peak below A which is due to a d$-$d
excitation \cite{Ellis08}. The d$-$d excitation also shows up in
ellipsometry data, though its spectral weight is suppressed.


In order to understand the observed discrepancy, it is useful to
consider the structural differences between the cuprate samples
studied here. Although the main structural building block in all
these samples is a plaquette composed of one Cu and four oxygens,
the network formed by these CuO$_4$ plaquettes is very different, as
shown in Fig.~\ref{fig:compare} insets. In Bi$_2$CuO$_4$, these
plaquettes are isolated, while they form an edge-sharing chain in
CuGeO$_3$. On the other hand, the plaquettes form a two-dimensional
corner-sharing network in the other three compounds. It is important
to realize that the interaction between the valence electron system
and the $1s$ core hole in the intermediate state is {\em non-local}
in the case of such corner-sharing copper oxides due to the 180
degree Cu-O-Cu bond in these compounds
\cite{vanVeenendaal93,Okada95}. Therefore, one can imagine that the
local core hole potential approximation used in Ref.~\cite{Ament07}
may break down in this case. The importance of non-locality can also
be inferred from the cluster calculation \cite{Li91}, in which a
cluster substantially larger than the plaquette was necessary to
explain the observed XAS spectrum in $\rm La_2CuO_4$.

We note that in their study of HgBa$_2$CuO$_{4+\delta}$, Lu et al.
reported that RIXS reveals more excitations than conventional
two-particle spectroscopy such as optical spectroscopy \cite{Lu05}.
In addition, RIXS experiments on La$_2$CuO$_4$ have shown that there
are many charge excitation peaks in addition to the lowest 2~eV peak
\cite{LCO-PRL,Collart06,Ellis08}. The observed fine structure, such
as the features B and D in Fig.~\ref{fig:detail}, thus could arise
from the difference in the matrix elements that enhances certain
spectral features selectively.

In summary, we show that overall spectral features of the {\em
indirect} resonant inelastic x-ray scattering response function are
in a reasonable agreement with the optical dielectric loss function
over a wide energy range. In the case of Bi$_2$CuO$_4$ and
CuGeO$_3$, we observe that the incident energy independent response
function, ${S}(\bf q=0,\omega)$, matches very well with the
dielectric loss function, Im(-1/$\varepsilon(\omega)$), suggesting
that the local core hole approximation treatment of
Ref.~\cite{Ament07} works well in these compounds. We also find that
corner-sharing two dimensional copper oxides exhibit more complex
excitation features than those observed in the dielectric loss
functions. Our study seems to suggest that one can probe
two-particle charge correlation function with indirect resonant
inelastic x-ray scattering.

\acknowledgements{We would like to thank Luuk Ament, Fiona Forte,
and J. van den Brink for the discussions. Research at the University
of Toronto was supported by the NSERC of Canada, Canadian Foundation
for Innovation, and Ontario Ministry of Research and Innovation.
Work at Brookhaven was supported by the U. S. DOE, Office of Science
Contract No. DE-AC02-98CH10886. Use of the Advanced Photon Source
was supported by the U. S. DOE, Office of Science, Office of Basic
Energy Sciences, under Contract No. W-31-109-ENG-38.}


\begin{thebibliography}{37}
\bibitem{pseudo}
\bibinfo{journal}{For $\rm Sr_2Cu_3O_4Cl_2$ and $\rm Sr_2CuO_2Cl_2$, the pseudo dielectric
functions were taken due to their planar surfaces.}
\bibitem{spin_part}
\bibinfo{journal}{We use the spinless fermion expression from Ref.~[3], since double occupancy can be ignored
in these copper oxide compounds due to the strong on-site Coulomb
interaction. Note also that the energy notation used here is
slightly different from that of Ref.~[3].}
\end{thebibliography}

\end{document}